# AUTOCORRELATION FUNCTIONS, COSMOLOGY AND INVESTIGATING THE CMB COLD SPOT WITH EMU-ASKAP RADIO CONTINUUM SURVEY

Syed Faisal ur Rahman[1,2]

[1] University of Karachi, Karachi, faisalrahman36@hotmail.com
[2] Institute of Business Management, Karachi

ABSTRACT

Galaxy angular-power spectrum and autocorrelation functions (ACFs) provide information about the distribution of matter by using galaxy counts as a proxy. In this study, we are going to estimate autocorrelation angular power spectrum and angular autocorrelation function for EMU-ASKAP 5 sigma sources and then compare them with results from NVSS. We will also use SUMSS data to compare ACF results using Landy-Szalay estimator. EMU-ASKAP will provide excellent opportunity to observe universe with high sensitivity and is likely going to observe millions of high redshift sources which will help in studying the clustering of the large scale structures, constraining cosmological parameters and exploring mysteries like the existence of a cosmic cold spot or the CMB cold spot as observed by both Planck and WMAP probes. We will discuss some possible ways, the CMB cold spot puzzle can be explored further by using the galaxy clustering, integral source count and galaxy bias analysis with a highly sensitive survey like EMU-ASKAP.

INTRODUCTION

Galaxy angular-power spectrum provides information about the distribution of matter by using galaxy counts as a proxy [1][2]. In this study, we are going to estimate autocorrelation angular power spectrum and angular autocorrelation function (ACF) for EMU 5 sigma sources [3][4][5][6] and then compare them with results from NVSS [1] [7]. We will also use SUMSS data to compare ACF results.

ANGULAR POWER SPECTRUM AND AUTOCORRELATION FUNCTION

We can write the angular power spectrum, $Cl_{gg}$ as [1]:

$$Cl_{gg} = 4\pi \int_{kmin}^{kmax} \frac{dk}{k} \Delta^2(k) \{Wl_g(k)\}^2 \qquad \text{Equation (1)}$$

Here $\Delta^2(k)$ is the logarithmic matter power spectrum, which can be calculated as:

$$\Delta^2(k) = \frac{k^3}{2\pi^2} P(k)$$

In this, $P(k)$ is the matter power spectrum. $Wl_g(k)$ is galaxy window function.

For this analysis we adopted Limber's approximation as discussed in [8][9]. This will give an expression for $Cl_{gg}$ as:

$$Cl_{gg} = 4\pi \int_0^{z*} \frac{dz}{c} \frac{H(z)}{X^2(z)} \{Wl_g(z)\}^2 P(k) \qquad \text{Equation (2)}$$

With k being approximated as [10]:

$$k = \frac{l + \frac{1}{2}}{\chi^2(z)}$$

With the window function defined as [9]:

$$Wl_g(z) = b(z)\frac{dN}{dz} + \frac{3\Omega_m}{2c}\frac{H_0^2}{H(z)}(1+z)X(z)\int_z^{z*} dz' \left(\frac{X(z') - X(z)}{X(z')}\right)(\alpha(z') - 1)\frac{dN}{dz'}$$

$$\text{Equation (3)}$$

The second part after '+' sign represents the contribution from the magnification bias [9][2]. It is dependent on $\alpha(z)$ which is the slope of the integral count ($N(>S) = CS^{-\alpha}$, see next section). In surveys like NVSS [1] [7] or SUMSS [11] where $\alpha$ is close to 1 [2], this part doesn't play a significant role. Magnification bias contribution also depends on the shape of $\frac{dN}{dz}$. We checked for EMU 5-sigma using SKADS data [12] and found it to play a negligible part in the overall galaxy autocorrelation power spectrum. However, with real data, it will be good to measure it again after carefully measuring $\alpha$ and $\frac{dN}{dz}$. In figure (1), we used SciPy's curve_fit [13] for the model, $N(z) = \frac{dN}{dz} = \left(\left(\frac{z}{a}\right)^b\right) exp\left(\frac{-cz}{a}\right)$ with a=0.40086143 b=0.60527401 c=0.54263201. Blue '+' represent SKADS data points, green solid line represents our model and others are fit from numpy's polyfit [14] function which are used for comparison here.

For EMU 5 sigma, galaxy bias b(z) is used as a weighted average of bias of different source types in each redshift bin and after z=3, we use a constant bias (see figure 2).

We use redshift distribution for NVSS as provided by [15] [16] and galaxy bias as discussed in [17]. For NVSS, we use a redshift dependent bias function [17]:
$$b(z) = 0.90[1 + 0.54(1+z)^2]$$
From 'Cl' values, we can get the angular correlation function (ACF) as [1]:

$w_{gg}(\theta) = \sum_{l=l_i}^{lmax} \frac{2l+1}{4\pi} Cl_{gg} Pl(cos\theta)$  Equation (4)

For our analysis we used $l_i$=2 and $l_{max}$=1000. This method of converting 'Cl' values into ACF values is suitable for both theory and large scale observations with large sky coverage (fsky). For small sky patches, a suitable estimator is Landy-Szalay estimator [18], which can be used as:

$w_{gg}(\theta) = \frac{DD(\theta) - 2DR(\theta) + RR(\theta)}{RR(\theta)}$  Equation (5)

Here,
DD=Two point count-count correlation from galaxy catalog.
DR=Two point cross correlation of counts from galaxy and random catalogs.
RR=Two point count-count correlation from random catalog.
Random catalog is developed in such a way that number of sources in random catalog are greater than or equal to ten times the number of sources in the galaxy catalog.

We use 100 bootstrap iterations to compute mean and variance for ACF estimates. For SUMSS, sources, we use $-65° < declination < -50°$ and $290° < right\ ascension < 340°$ with flux>20 mJy. For NVSS's flux > 10 mJy configuration, we use $5° < declination < 58°$ and $12° < right\ ascension < 34°$.

We also removed sources > 1Jy from our analysis. We use WMAP 9 years cosmological parameter results to perform theoretical calculations [19]. We can see angular power spectrum results in figures (3 and 4) where we can see comparison between EMU 5 sigma estimates with theoretical values of NVSS and Blake 2004 results.

Our ACF results can be seen in figures (5 and 6) where EMU 5 sigma results are compared with NVSS flux > 10 mJy, SUMSS > 20 mJy and NVSS ACF power law fit results from [20].

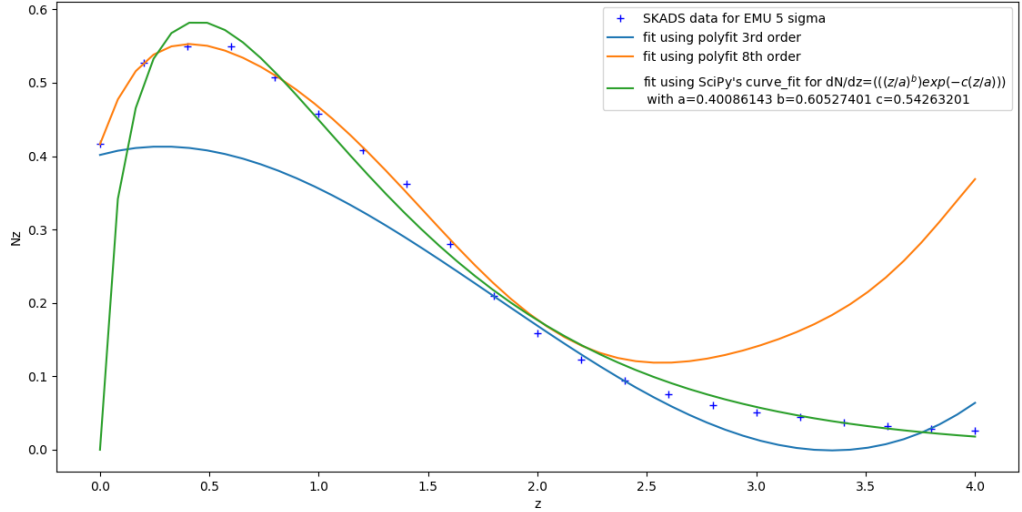

**Figure 1: $\frac{dN}{dz}$ from SKADS for EMU 5 sigma.**

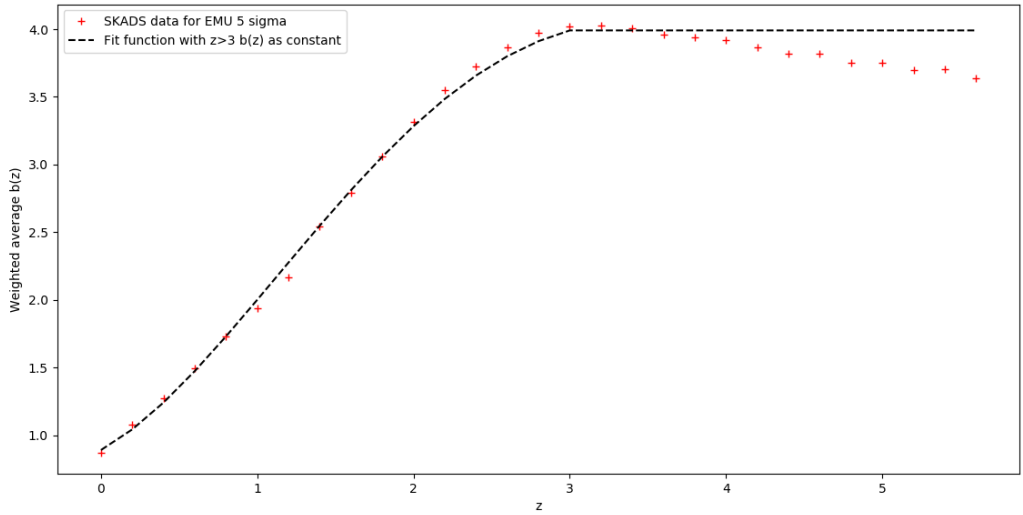

**Figure 2: linear weighted average bias b(z) for EMU 5 sigma sources from skads and the best fit function. In the fit function, we use a constant value after z=3. Here, weight is based on the proportion of source types present in each redshift bin.**

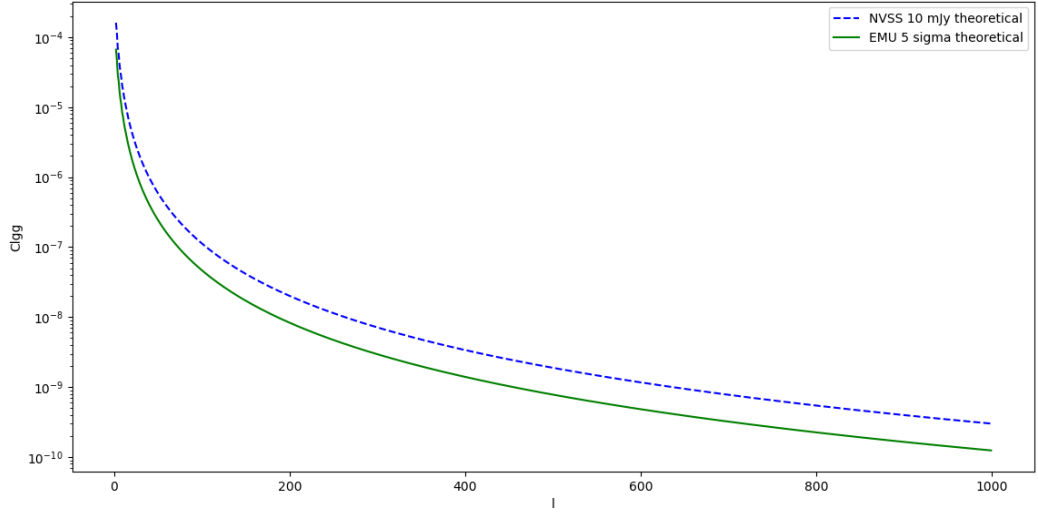

**Figure 3: Comparison of autocorrelation angular power spectrum between theoretical estimates under Λ-CDM assumption for EMU 5 sigma and NVSS.**

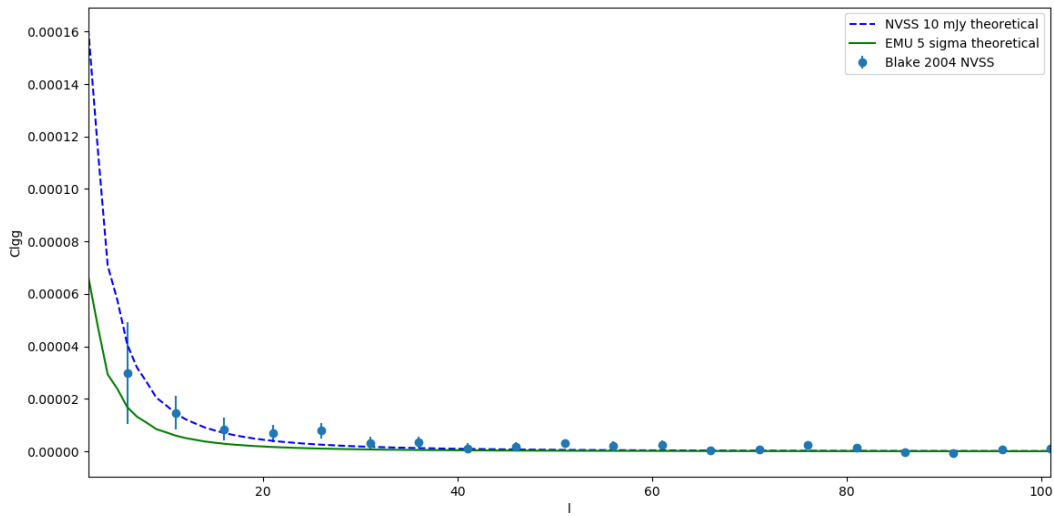

**Figure 4: Comparison between theoretical estimates for EMU 5 Sigma and NVSS with Blake 2004 values.**

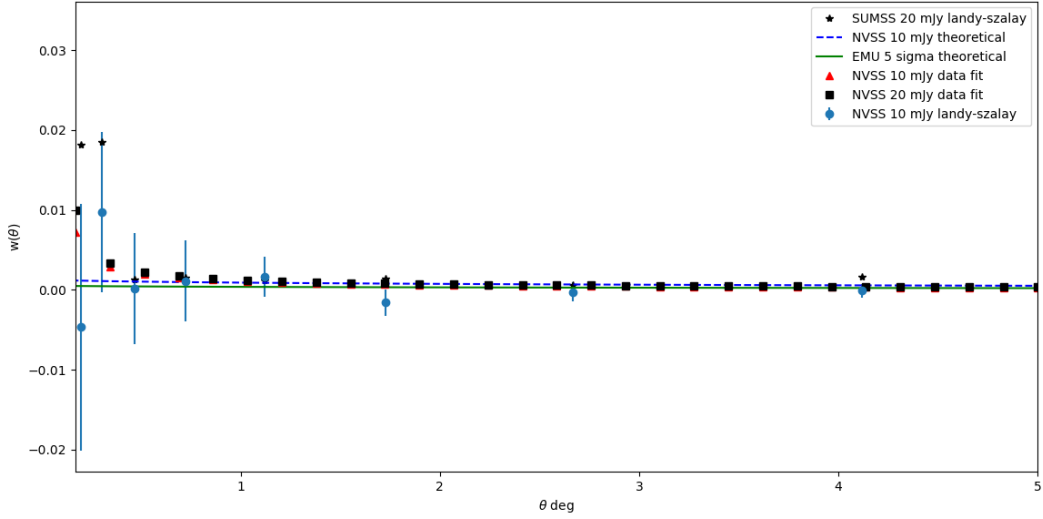

**Figure 5:** Comparison of autocorrelation function (ACF)'s from EMU 5 sigma, NVSS theory, NVSS power law fits from Blake 2002, and NVSS and SUMSS ACF's measured using AstroML's Landy-Szalay estimator.

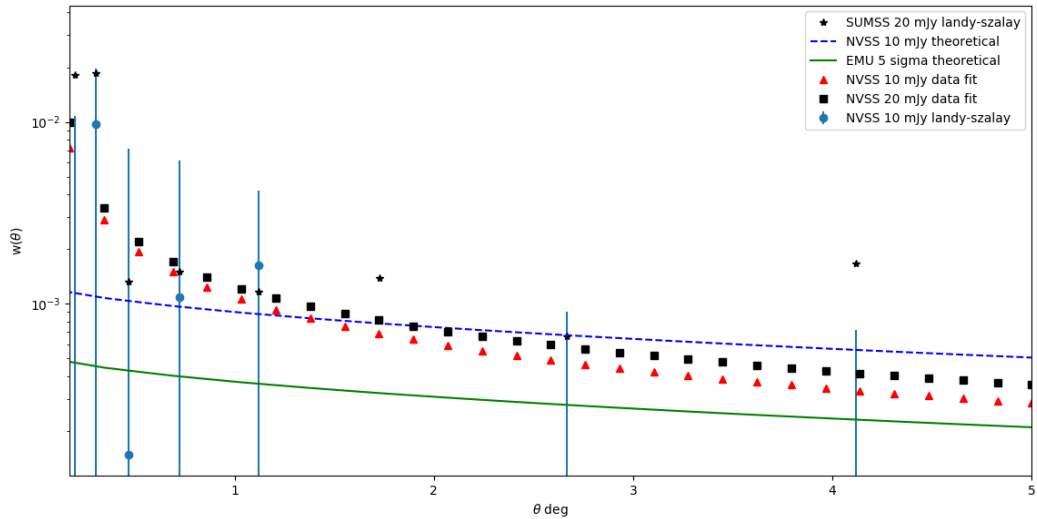

**Figure 6:** Comparison of autocorrelation function (ACF)'s from EMU 5 sigma, NVSS theory, NVSS power law fits from Blake 2002, and NVSS and SUMSS ACF's measured using AstroML's Landy-Szalay estimator. Here, the results are scaled to focus on lower theta contribution.

DIFFERENTIAL AND INTEGRAL SOURCE COUNT.

In order to understand the distribution of sources and estimate confusion, magnification bias and other useful quantities, differential and integral source counts provide useful means (Condon 2007, Rahman & Iqbal 2019). Theoretically, we can calculate integral source counts from differential count power law distributions as [21] [5]:

$$N(S>S_{min}) = \int n(S) dS \quad \text{Equation (6)}$$

Here n(S) is the differential source count power law probability distribution in $Jy^{-1} Sr^{-1}$ as [21] [5]:

$$n(S) = \frac{dN}{dS} = kS^{-\gamma} \quad \text{Equation (7)}$$

For EMU-5 sigma sources using SKADS, we get an estimate of $k \approx 57.24$ and $\gamma \approx 2.18$ [4] [5].

Smin in equation (6) is can be the rms or any suitable the lower bound value based on the survey science goals and technical specifications.

Integral source counts from observational data are usually are used to fit a power law curve in the form of:

$$N(S) = CS^{-\alpha}$$  Equation (8)

This gives us 'α' which is not only useful in measuring the magnification bias part of equation (3). Using SKADS database, we get α ≈ 1.18 an C ≈ 48.5 [4] [5].

In order to perform statistical error analysis for galaxy continuum surveys, the shot-noise measurements play an important part. Shot noise estimates or measurements are required to calculate the signal to noise ratios, measure error bars, and obtain correct covariance matrices, especially in relation to the theoretical or observed 'Cl' values obtained during cross or autocorrelation studies.

We can define shot-noise as

$$Shot\ Noise = \frac{\Delta\Omega}{N}$$

Where ΔΩ =observed area of the survey in steradian and N=number of sources observed in the total survey area. Shot-noise can be calculated from the number count per steradian (Ns), by using the simple relation:

$$Shot\ Noise = 1/Ns$$

Where Ns is the number of sources per steradian. Ns can be calculated using equation (8) by using the differential source count power law fit from equation (7).

# EMU-INVESTIGATING THE CMB COLD SPOT

An interesting phenomenon related to the cosmic microwave background (CMB) is the presence of an unusal cold spot. CMB Cold Spot is an unusually cold region on CMB maps observed by both WMAP and Planck. [22][23] This area temperature deviates from average temp by >70 micro K. The area is located at RA 03h 15m 05s, DEC −19° 35′ 02″. With EMU-ASKAP's unprecedented depth and high resolution, we are planning to investigate this region. We will use EMU data to investigate if unusual source clustering, source properties or any other unusual phenomenon is causing this region to have an unusually low temperature. We will also use data from other surveys like NVSS, 2MASS, WISE,2CSz [22][23] and other surveys to explore this region. We would also like to check its influence on cosmological parameter fitting particularly in relation with curvature and Hubble Constant problems. Getting redshift information and classification will be important (EMU has separate working groups which are working on redshift estimates. Their work will be helpful). We can test a lot of things using the cold spot area and compare results with other areas (ACF, CCF, redshift distributions,galaxy bias evolution,power laws for integral and differential counts etc.) [22][23].

Some of the things we are planning to test are (Rahman 2020a, Rahman 2020b):

- Integrated Sachs Wolfe Effect: EMU Cosmology group is working for the whole survey area and we can use this to dig into the cold spot problem [24] [25][4].
- Unusual clustering, masses, absence or lower than usual presence of dark matter and bias evolution in the CMB cold spot region galaxies (Related to both Auto and Cross correlations) [22][23].
- Cosmological modeling: Flat-Lambda CDM, dynamic dark energy EoS, curvature issue and alternates to standard cosmological model etc. We can compare measurements from the CMB cold spot with other regions and see if there are deviations and then look for possible reasons [22][23][26].
- Unusual structures: Possible features similar to "Great Attractor" or "Dipole Repeller" or "South Pole Wall" or something else and how? [22][23]
- Unusual power law fits (Integral and differential). Magnification bias for Clgg will be difficult but we can separately calculate alpha from integral source counts and gamma from differential to see unusual clustering from other patches of sky. Clustering anomaly can indicate unusual masses [4][5][22][23].

# CONCLUSION

In this study, we estimated autocorrelation angular power spectrum and angular autocorrelation function (ACF) for EMU 5 sigma sources and then compared our results with results from NVSS. We then used SUMSS data to compare ACF results. We also discussed differential and integral source counts, and their role in estimating various useful quantities like the magnification bias which are useful in modeling angular power spectrum of galaxies and ACFs. We then discussed CMB Cold Spot and our strategy to explore and understand this phenomenon using the EMU-ASKAP survey.

# ANCKNOWLEDGMENT: